\documentstyle[12pt,psfig]{article}
\begin{document}
\begin{titlepage}

\pagestyle{empty}
\begin{flushright}
{\footnotesize UFRN-DFTE\\
GAC-99040\\
July 1999}
\end{flushright}

\vskip 1.0cm

\begin{center}
{\large \bf Minimal Angular Size of Distant Sources in Open, $\Lambda$CDM, and Scalar Field Cosmologies}
\vskip 1cm
\renewcommand{\thefootnote}{\alph{footnote}}
J. A. S. Lima$^{}$\footnote{e-mail:limajas@dfte.ufrn.br}
and J. S. Alcaniz$^{}$\footnote{e-mail:
alcaniz@dfte.ufrn.br}
\end{center}
\vskip 0.5cm

\begin{quote}
\begin{center}
{\small $^{}$Departamento de F\'{\i}sica Te\'orica e Experimental\\
Universidade
Federal do Rio Grande do Norte\\
59072-970 Natal - RN - Brazil}
\end{center}
\end{quote}

\vskip 1.5cm

%\documentstyle[12pt,psfig]{article}
%\documentstyle[prd,aps,preprint,psfig]{revtex}
%\begin{document}
%%\draft
%\title{\bf Minimal Angular Size of Distant Sources in Open, $\Lambda$CDM, and Scalar Field Cosmologies}
%
%\maketitle
%\author{J. A. S. Lima\footnote[1]
%{limajas@dfte.ufrn.br}  and J. S. Alcaniz\footnote[2]
%{alcaniz@dfte.ufrn.br}}
%Universidade
%Federal do
%Rio Grande do Norte,
%\\Departamento de F\'{\i}sica,
%Caixa Postal 1641, \\59072-970 Natal, RN,
%Brazil
%\date{march 06, 1999}%5
%
%\smallskip
%%\address{
%%Universidade
%%Federal do
%%Rio Grande do Norte,
%%\\Departamento de F\'{\i}sica,
%Caixa Postal 1641, \\59072-970 Natal, RN,
%Brazil
%\date{march 06, 1999}
%\maketitle
%
\begin{abstract}

We propose a simple method for determining the redshift
$z_{m}$ at which the angular size of an extragalactic source
with
fixed proper diameter takes its minimal value. A closed
analytical
expression, which is quite convenient for numerical evaluation
is derived. The method is exemplified with the following FRW
type expanding
universes:  the open matter dominated models ($\Omega_{\Lambda}
= 0$),
a critical density model with cosmological constant
($\Omega_{\Lambda} \neq 0$), and the class of scalar field
cosmologies proposed by Ratra and Peebles. The influence of
systematic evolutionary effects is briefly discussed.
\end{abstract}
\end{titlepage}
\newpage

\section{Introduction}

The angular size - redshift relation, $\Theta (z)$, is a
kinematic test which potentially may discriminate the several
cosmological models proposed in the literature. As widely
known,
because of the spacetime curvature, the expanding universe acts
gravitationally
as a lens of large focal length. Though nearby objects are not
affected, a fixed angular size of an extragalactic source is
initially seen decreasing up to a minimal value, say, at a
critical redshift
($z_m$), after which increasing for higher redshifts. The
precise
determination of $z_m$, or equivalently, the corresponding
minimal
angular size value $\Theta(z_m)$, may constitute a powerful
tool in
the search for deciding which are the more realistic world
models. This
lensing effect was first predicted by Hoyle, originally  aiming
to distinguish
between the steady-state and Einstein-de Sitter cosmologies \cite{hoyle}.
Later on, the accumulated evidences against the steady state
(mainly  from
CMBR) have put it aside, and more recently, the same is
occurring with the
theoretically favoured critical density FRW model \cite{krauss2,Riess,Perlmutter,Alcaniz,Roos}.

The data concerning the angular size - redshift
relation are until nowadays somewhat controversial, specially
because
they envolve at least two kinds of observational dificulties.
First, any
large redshift object may have a wide range of proper sizes,
and, second,
evolutionary and selection effects probably are not negligible.
The $\Theta(z)$  relation for some extended
sources samples seems to be quite imcompatible with the
predictions of the  standard FRW model when the latter effects
are not taken into account \cite{Sand,Kapa1,Kapa2}.
There have also been some claims that
the best fit model for the observed distribution of high
redshifts extended
objects is provided by the standard Einstein-de
Sitter  universe ($q_o={1 \over 2}$, $\Omega_\Lambda=0$) with
no significant  evolution \cite{Bucha}.
Parenthetically, these results are in contradiction with
recent observations
from type Ia supernovae, which seems to ruled out world models
filled only by baryonic matter, and more generally,
any model with positive deceleration parameter \cite{Riess,Perlmutter}. The same happens with
the corresponding bounds using the ages of old high redshift
galaxies \cite{Alcaniz,Dunlop,Krauss3}.

The case for compact radio sources is also of great interest.
These objects
are apparently less sensitive to evolutionary effects since
they are
short-lived ($\sim 10^{3} yr$) and much smaller than their host
galaxy. Initially, the
data from a sample of 82  objects gave  remarkable suport for
the Einstein-de Sitter Universe \cite{Kell}. However,
some analysis suggest that Kellerman has not really detected a
significant increasing beyond the minimum \cite{Dabro,Step,Cool}. Some
authors have also argued that models where $\Theta(z)$
diminishes and after a given $z$ remains constant may also
provide a good fit to Kellerman's data. In particular, by
analysing a subset of 59 compact sources within the same
sample,
Dabrowski et al. (1995) found that no useful bounds on the
value of the
deceleration parameter $q_o$ can be derived. Further, even
considering that
Euclidean angular sizes ($\Theta \sim z^{-1}$) are excluded at
99$\%$
confidence level, and that the data are consistent with
$q_o=1/2$, they
apparently do not rule out extreme values of the deceleration
parameter as
$q_{o} \sim 5$ \cite{Step}. More recently, based
in a more
complete sample of data, which include the ones originally
obtained by
Kellermann, it was argued that the $\Theta(z)$ relation may be
consistent with any model of the FRW class with deceleration
parameter $\leq 0.5$ \cite{Gurv}.

In this context, we discuss here how the critical redshift
giving the turn-up in angular sizes is determined for any
expanding cosmology based on the FRW geometry. An analytical
expression quite convenient for numerical evaluation is
derived. The approach is exemplified for three different
models of current cosmological interest: (i) open matter
dominated FRW universe (OCDM), (ii) flat FRW type models with
cosmological
constant ($\Lambda$CDM), (iii) the class of scalar field
cosmologies (SF)
proposed by Ratra and Peebles \cite{Ratra}. Hopefully, the results
derived
here may be useful near future, when more accurate data become
available.

\section{The method}

Let us now consider the FRW line element $(c=1)$
\begin{equation}
 ds^2 = dt^2 - R^{2}(t) [d\chi^{2} + S^{2}_{k}(\chi) (d
\theta^2 +
\rm{sin}^{2} \theta d \phi^{2})]   \quad  ,
\end{equation}
where $\chi$, $\theta$, and $\phi$ are dimensionless comoving
coordinates,  $R(t)$ is the scale factor, and $S_{k}(\chi)$
depends on
the curvature parameter ($k=0$, $\pm 1$). The later function is
defined
by one of the following forms: $S_k (\chi) = \rm{sinh} (\chi)$,
$\chi$,
$\rm{sin} \chi$, respectively, for open, flat  and closed
Universes.

In this background, the angular size-redshift relation for a
rod of intrinsic length $D$ is easily obtained by integrating
the spatial part of the above expression for $\chi$ and $\phi$
fixed. One finds
\begin{equation}
\theta(z) = {D (1 + z) \over R_{o}S_{k}(\chi)}  \quad  .
\end{equation}
The dimensionless coordinate $\chi$ is given by
\begin{equation}
\chi(z) = {1 \over H_o R_o} \int_{(1 + z)^{-1}}^{1} {dx \over x
E(x)}
 \quad  ,
\end{equation}
where $x = {R(t) \over R_o} = (1 + z)^{-1}$ is a convenient
integration variable.
For the three kinds of cosmological models considered here
(OCDM, $\Lambda$CDM and SF) the
dimensionless function $E(x)$ assume one of the following
forms:
\begin{equation}
E_{FRW}(x) = \left[1 - \Omega_{M} + \Omega_{M}
x^{-1}\right]^{{1
\over 2}} \quad  ,
\end{equation}
\begin{equation}
E_{\Lambda}(x) = \left[(1 - \Omega_{\Lambda}) x^{-1} +
\Omega_{\Lambda}x^{2}\right]^{{1 \over 2}} \quad  ,
\end{equation}
\begin{equation}
E_{SF}(x) = \left[(1 - \Omega_{\phi})x^{-1} +
\Omega_{\phi}x^{{4-\alpha}
\over {2 + \alpha}}\right]^{{1}\over{2}}  \quad  ,
\end{equation}
where $\Omega_{M} = {{8 \pi G\rho_{M}} \over 3 H_{o}^{2}}$,
$\Omega_{\Lambda} = {\Lambda \over 3H_{o}^{2}}$ and
$\Omega_{\phi} =
{8 \pi G\rho_{\phi} \over 3 H_{o}^{2}}$, are the present day
density
parameters associated with the matter component, cosmological
constant and the scalar field $\phi$, respectively. Notice that
equations (5) and (6) become identical if one takes $\alpha =
0$ in
the later, thereby showing that the scalar field model proposed
by
Ratra and Peebles may kinematically be equivalent to a flat
$\Lambda$CDM cosmology.

The redshift $z_{m}$ at which the angular size takes the
minimum
value is the one cancelling out the derivative of $\Theta$ with
respect to $z$.
Hence, from (2) we have the condition
\begin{equation}
S_k (\chi_m) = (1 + z_m)S'_k (\chi_m)  \quad  ,
\end{equation}
where $S'_k (\chi) = {\partial S_{k} \over \partial
\chi}{\partial
\chi \over \partial z}$, a prime denotes differentiation with
respect
to $z$ and by definition $\chi_{m}= \chi(z_{m})$. To proceed
further,
observe that (3) can readily be differentiated yielding,
respectively, for the standard FRW (matter dominated),
$\Lambda$CDM
and scalar field cosmologies
\begin{equation}
(1 + z_{m})\chi'_{m} = {(R_o H_o)^{-1} \over \left[1 - \Omega_M
+
\Omega_M (1 + z_m)\right]^{{1 \over 2}}} = (R_o H_o)^{-1}
F(\Omega_{M}, z_m) \quad  ,
\end{equation}
\begin{equation}
(1 + z_{m})\chi'_{m} = {(R_o H_o)^{-1} \over \left[(1 -
\Omega_{\Lambda})(1 + z_m) + \Omega_{\Lambda}(1 +
z_m)^{-2}\right]^{{1 \over 2}}} = (R_o H_o)^{-1}
L(\Omega_{\Lambda},
z_m) \quad  ,
\end{equation}
\begin{equation}
(1 + z_{m})\chi'_{m} = {(R_o H_o)^{-1} \over \left[(1 -
\Omega_{\phi})(1 + z_m) + \Omega_{\phi}(1 + z_m)^{{\alpha - 4
\over
\alpha + 2}}\right]^{{1 \over 2}}} = (R_o H_o)^{-1}
S(\Omega_{\phi}, \alpha, z_m) \quad  .
\end{equation}

Now, inserting the above equations into (7) we find
for the cases above considered
\begin{equation}
{1 \over (1 - \Omega_{M})^{1 \over 2}}{\rm{tanh}}\left[(1 -
\Omega_{M})^{1 \over 2}\int_{(1 + z_m)^{-1}}^{1} {dx \over x
E_{FRW}(x)}\right] =
F(\Omega_{M}, z_m)  \quad  ,
\end{equation}
\begin{equation}
\int_{(1 + z_m)^{-1}}^{1} {dx \over x E_{\Lambda}(x)} =
L(\Omega_{\Lambda}, z_m)  \quad  ,
\end{equation}
\begin{equation}
\int_{(1 + z_m)^{-1}}^{1} {dx \over x E_{SF}(x)} =
S(\Omega_{\phi},
\alpha, z_m)  \quad  .
\end{equation}

The meaning of equations (11)-(13) is self evident. Each one
represents an  integro-algebraic equation for the critical
redshift
$z_m$ as a function of the physically meaningful parameters of
the
models. In general, these equations cannot be solved in closed
analytical form for $z_m$. However, as one may  check, if we
take the
limit $\Omega_M \rightarrow 1$ in (11), the value $z_m = {5
\over 4}$
is readily achieved, which corresponds to the well known
standard
result for the dust filled FRW flat universe. The interesting
point is
that expressions (11)-(13) are quite convenient for numerical
evaluations. As a matter of fact,
their solutions can straightforwardly be obtained, for
instance, by
programming the integrations using simple numerical recipes in
FORTRAN.

In Fig.~1 we show the diagrams of $z_m$ as a function of the
density parameter for each kind of model. As expected, in the
standard FRW model, the critical redshift starts at $z_m=1.25$
when $\Omega_M$ goes to unity. This value is pushed to the
right direction, that is, it is
displaced to higher redshifts as the $\Omega_M$ parameter is
decreased. For instance, for $\Omega_M=0.5$ and $\Omega_M=0.2$,
we find $z_m=1.58$ and $z_m=2.20$, respectively. In the
limiting case, $\Omega_M \rightarrow 0$, there is
no minimum at all since $z_{m} \to \infty$. This means that the
angular size decreases monotonically as a function of the
redshift.
For the scalar field case, one needs to fix the value of
$\alpha$ in
order to have a bidimensional plot. Given a value of
$\Omega_{\phi}$,
the minimum is also displaced for higher redshifts when the $\alpha$
parameter diminishes. Conversely, for a fixed value of
$\alpha$, the
minimum moves for lower redshifts when $\Omega_\phi$ is
decreased. The limiting case ($\alpha=0$) is fully equivalent
to a
$\Lambda$CDM model. As happens in the limiting case $\Omega_M
\rightarrow 0$ ($\Omega_{\Lambda} = 0$), the  minimal value for
$\Theta(z)$
disappears when the cosmological constant contributes all the
energy density
of the Universe, that is, $z_m \rightarrow \infty$  if
$\Omega_M \rightarrow 0$
and $\Omega_{\Lambda} \rightarrow 1$ (in this connection see
also \cite{krauss1}). For the class of models considered in this
paper, the
redshifts having the minimal angular size are displayed for
several values of $\Omega_M$ and $\alpha$ in Table 1. As can be
seen there, the critical
redshift at which the angular size is a minimal cannot alone
discriminate
between world models since different scenarios may provide the
same $z_m$
value. However, when combinated  with other tests, some
interesting
constraints on the cosmological models can be  obtained.
For example, when $\Omega_\phi$ is bigger than $0.55$, the model proposed by Ratra and  Peebles yields a $z_m$
between the standard FRW flat model and the  $\Lambda$CDM cosmology. Then,
suposing that the universe is really accelerating today ($q_{o} < 0$), as
indicated recently by measurements  using type Ia supernovae \cite{Riess,Perlmutter}, and by considering
the results by Gurvits et al. \cite{Gurv}, i.e., that the data are
compatible with
$q_{o} \leq 0.5$, the Ratra and Peebles models with $0 < \alpha
\leq 4$ seems
to be more in accordance with the angular size data for compact
radio sources
than the $\Lambda$CDM model.

\begin{table}[t]
\begin{center}
\begin{tabular}{rrlll}
\hline
\multicolumn{1}{c}{$\Omega_{m}$ ($z_{m}$)}&
\multicolumn{1}{c}{$\Omega_{\Lambda}$ ($z_{m}$)}&
\multicolumn{1}{c}{$\Omega_{\phi} (\alpha = 2)$ ($z_{m}$)}&
\multicolumn{1}{c}{$\Omega_{\phi} (\alpha = 4)$ ($z_{m}$)}&
\multicolumn{1}{c}{$\Omega_{\phi} (\alpha = 6)$ ($z_{m}$)}\\
\\
\hline
1.0 (1.25)& 1.0 ($\infty$)& 1.0 (2.16)& 1.0 (1.72)& 1.0
(1.57)\\
\\
0.8 (1.35)& 0.8 (1.76)& 0.8 (1.65)& 0.8 (1.53)& 0.8 (1.46)\\
\\
0.7 (1.41)& 0.7 (1.60)& 0.7 (1.55)& 0.7 (1.47)& 0.7 (1.42)\\
\\
0.5 (1.58)& 0.5 (1.44)& 0.5 (1.42)& 0.5 (1.38)& 0.5 (1.36)\\
\\
0.2 (2.20)& 0.2 (1.31)& 0.2 (1.30)& 0.2 (1.30)& 0.2 (1.29)\\
\hline
\end{tabular}
\caption{Critical redshift $Z_m$ in OCDM, $\Lambda$CDM, and scalar field cosmologies for some selected values of the density parameters.}
\end{center}
\end{table}
\pagebreak

\begin{figure}
\vspace{.2in}
\centerline{\psfig{figure=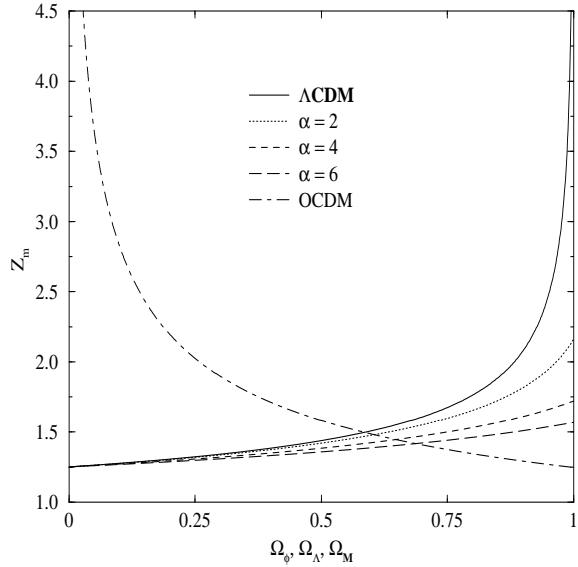,width=3truein,height=3truein}
\hskip 0.1in}
\caption{Critical redshift $Z_m$ as a function of the density
parameter in open, ${\Lambda}\rm{CDM}$ and scalar field cosmologies. Solid
curve is the prediction for a model with nonnull cosmological constant. The same curve is also obtained as a limiting case ($\alpha \rightarrow 0$) of the scalar field cosmology  proposed by Ratra and Peebles}
%\label{flt}
\end{figure}

It is worth notice that the same procedure may be applied when
evolutionary
and/or selection effects due to a linear size-redshift or to a
linear size-luminosity dependence are taken into account. As
widely believed, a plausible way of standing for such effects
is to consider that the intrinsic linear size has a similar
dependence on the redshift as the coordinate dependence, i.e.,
$D = D_o (1 + z)^{c}$, being $c < 0$ (see, for instance,
\cite{Bucha} and Refs. therein). In this case,
equations (11)-(13) are still valid but the functions
$F(\Omega_{M}, z_m)$, $L(\Omega_{\Lambda}, z_m)$, and
$S(\Omega_{\phi}, \alpha, z_m)$ must be divided by a factor $(1
+ c)$. The displacement of $z_{m}$ relative to the case with no evolution
($c = 0$) due to the effects cited above may be unexpectedly
large. For example, if one takes $c = -0.8$ as found by
Buchalter et al. \cite{Bucha}, the redshift of the minimum angular
size for the Einstein-de Sitter case ($\Omega_{M} = 1$) moves
from $z_{m} = 1.25$ to $z_{m} = 11.25$. In particular, this
explains why the data of Gurvits et al. \cite{Gurv}, although
apparently in agreement with the Einstein-de Sitter universe,
do not show clear evidence for a minimal angular size close to
$z = 1.25$, as should be expected for this model.

\hspace{0.3cm}
{\bf Acknowledgments:}This work was partially
supported by the project Pronex/FINEP (No. 41.96.0908.00) and
Conselho Nacional de Desenvolvimento Cient\'{\i}fico e
Tecnol\'ogico - CNPq (Brazilian Research Agency).

\end{document}